# ULTRA-LOW SURFACE RESISTANCE VIA VACUUM HEAT TREATMENT OF SUPERCONDUCTING RADIOFREQUENCY CAVITIES

S. Posen*, A. Romanenko, A. Grassellino, O. S. Melnychuk, D. A. Sergatskov

*Fermi National Accelerator Laboratory, Batavia, Illinois, 60510, USA*

We report on an effort to improve the performance of superconducting radiofrequency cavities by the use of heat treatment in a temperature range sufficient to dissociate the natural surface oxide. We find that the residual resistance is significantly decreased, and we find an unexpected reduction in the BCS resistance. Together these result in extremely high quality factor values at relatively large accelerating fields $E_{acc}$ ~20 MV/m: $Q_0$ of 3-4×10$^{11}$ at <1.5 K and $Q_0$ ~5×10$^{10}$ at 2.0 K. In one cavity, measurements of surface resistance versus temperature showed an extremely small residual resistance of just 0.63 ± 0.06 nΩ at 16 MV/m. SIMS measurements confirm that the oxide was significantly dissociated, but they also show the presence of nitrogen after heat treatment. We also present studies of surface oxidation via exposure to air and to water, as well as the effects of very light surface removal via HF rinse. The possibilities for applications and the planned future development are discussed.

## I. INTRODUCTION

Superconducting radiofrequency (SRF) cavities are widely used in modern particle accelerator facilities (for examples of recent and near-future large SRF accelerators, see Refs. [1]–[6]) to transfer energy to beams of charged particles. By treating the niobium walls of the cavities with state-of-the-art surface processing techniques and cooling them to ~2 K, it is possible to reach accelerating fields $E_{acc}$ in the scale of tens of MV/m while maintaining a quality factor $Q_0$ in the 10$^{10}$ range.

As progress in R&D pushes $Q_0$ and $E_{acc}$ higher, a given investment in an accelerator facility can result in beam parameters—such as energy and duty factor—that are improved compared to what would have been possible with the previous state-of-the-art. To be realizable in applications, new treatments must not just achieve high $Q_0$ and $E_{acc}$ values, but must do so with high reproducibility. Nitrogen infusion is an example of a treatment that shows extremely promising $Q_0$ and $E_{acc}$ performance in several tests [7], but has had limited reproducibility in some laboratories [8], [9]. The process starts with a high temperature heat treatment of a cavity in a vacuum furnace at ~800 C, which is expected to dissolve the natural surface oxide layer. The furnace temperature is then lowered to 100-200 C, and low pressure nitrogen gas is injected to the furnace. After several hours of the low temperature nitrogen treatment on the post-oxide dissolution surface, SIMS (secondary ion mass spectrometry) measurements show that nitrogen interstitials can be found several tens of nm deep [7]. Unlike most other state-of-the-art cavity preparation methods, no material removal (e.g. via electropolishing (EP)) is done after furnace treatment. This likely makes it extremely sensitive to the pumping system and cleanliness of the furnace in which the infusion is performed.

To try to improve reliability, we developed a method to perform a nitrogen infusion-like treatment of a cavity *after* it had been assembled to vacuum hardware in a cleanroom environment. During the heat treatment, the inner cavity surface is expected to be exposed only to carefully cleaned and assembled surfaces, rather than to a vacuum furnace. The method has the added benefit of making it possible to measure cavity performance after the oxide dissolution step but before the surface oxide regrows. Several initial tests were performed on cavities after only the oxide dissolution step, and before adding a low temperature step with nitrogen. The results of these tests, as well as one with the addition of the low temperature nitrogen step, are presented in this paper.

In this paper, we will frequently refer to the surface resistance $R_s$ when discussing how treatment modifies $Q_0$. $R_s$ is related[1] to $Q_0$ via a geometric factor $G$ through $R_s = G/Q_0$. $R_s$ is, at temperatures near the critical temperature $T_c$, generally determined by the intrinsic properties of the superconductor ($T_c$, coherence length $\xi$, etc.) in a strongly temperature-dependent component of the surface resistance based on the BCS theory of superconductivity (referred to as the "BCS resistance" $R_{BCS}$) [10]–[12]. The BCS resistance approaches zero for temperature $T<<T_c$, and sources of surface resistance that are not intrinsic to the superconductor can begin to dominate, usually referred to as the residual resistance $R_{res}$ ($R_s = R_{BCS} + R_{res}$). The residual resistance is known to be increased by extrinsic factors such as trapped flux and electron loading (e.g. field emission, multipacting), but an "intrinsic" component of the residual resistance is often discussed as well, with little understanding of its origin or expected value [10]. We will show that the treatment methods described in this paper will modify $R_{BCS}$ and $R_{res}$ in different ways, and may help to set an upper bound on the contribution of the "intrinsic" $R_{res}$.

---

[1] This relation is approximate, but is expected to be a satisfactory approximation for the study presented here

*Email: sposen@fnal.gov

## II. EXPERIMENTAL PROCEDURE

For this study, several 1.3 GHz single cell cavities made from bulk niobium were prepared with established SRF treatment methods (electropolishing (EP), 75/120 C bake, nitrogen doping) as a baseline. As is typical, the cavities were ultrasonically cleaned, high pressure water rinsed (HPR), and assembled to vacuum hardware, including RF feedthroughs, a burst disc, and a pumpout port closed by a right angle valve. The assembled cavities were placed in a vertical test dewar and RF tested to establish baseline $Q_0$ vs $E_{acc}$ performance.

In the baseline test, the inner and outer surfaces of the cavities would be expected to be covered by the natural oxide, which is composed primarily of $Nb_2O_5$. Surface studies of niobium samples have shown (e.g. via XPS [13], [14] or X-ray reflectivity [15]) that this oxide can be dissociated by vacuum heat treatment at temperatures above ~300 C into Nb, O, and a small amount of residual suboxides.

After the heat treatment, a new oxide layer will grow if the surface is then exposed to air or water, so each of the cavities was heat treated while still assembled to its vacuum hardware. After the baseline measurements, an assembled cavity would be transported to an oven, which uses forced convection of heated air to maintain an internal temperature of up to ~300 C. All of the vacuum hardware was compatible with this temperature range except for the AlMg gaskets used to seal the cavities, so water cooling was implemented to lower the temperature of the flanges to an acceptable temperature. Heating bands around the beampipes of the cavity were used to offset the water cooling so that the RF volume of the cavity was maintained in the >300 C range, as monitored by thermocouples. To prevent absorption of atmospheric gases (including oxygen and hydrogen) in its exterior surface, the cavity was installed in the oven inside a steel can, into which argon gas was purged at a constant rate. The argon atmosphere was sampled into an oxygen monitor, which consistently read <1000 ppm. The apparatus is shown in Figure 1.

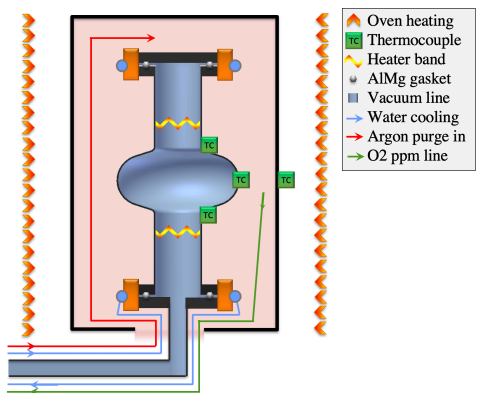

*Figure 1: Apparatus described in the text used to heat treat the niobium surrounding the elliptically-shaped RF volume at a temperature of >300 C. Precautions were taken to keep the AlMg gaskets cool and to prevent uptake of atmospheric gases in the exterior of the cavity wall.*

To carry out the heat treatment, the cavity temperature was ramped from room temperature at a rate of approximately 2 C/min to the maximum temperature for approximately 2.5 hours before turning off the heating power. Only a limited number of thermocouples were placed around the cavity, but the temperature range over the surface is expected to have been in the range 250-400 C. Because this temperature range is between that used in the ~120 C low temperature bake used to prevent High Field Q-Slope (HFQS) [16], [17] and the >600 C high temperature furnace treatment used to degas hydrogen, we will refer to this treatment as a "medium temperature bake" or "mid-T bake." During the mid-T bake, the cavity internal volume would be pumped with a turbomolecular pump. For most of the cavities presented here, after the mid-T bake, the oven would be cooled down to room temperature. In the case of one cavity, TE1AES011, the mid-T bake was maintained at temperature for 22 hours, and, before the cooldown, the argon purge was stopped and the oven temperature was lowered to 120 C. The cavity was backfilled with 25 mtorr of nitrogen via a needle valve (with the turbopump off) and this was maintained for 48 hours. This was done to mimic a 120 C nitrogen infusion.

A residual gas analyzer on the pump cart was used to leak check the cavities before and after the heat treatment via spraying helium on vacuum connections. A leak was discovered in the vacuum line connected to the burst disc on TE1AES012 after it received heat treatment, suggesting that a leak opened during the treatment. It was a relatively small leak ~$1 \times 10^{-9}$ torrL/s, and tightening the bolts on the flange fixed the leak.

After mid-T bake, the cavities were transported to the vertical dewar for a second cryogenic RF test. None of the cavities were disassembled or vented between the baseline RF test and this second RF test. However, after the second RF test, TE1PAV008 and TE1AES012 were vented with nitrogen gas after vertical test, partially disassembled, then given a high pressure water rinse before reassembly and a third RF test. This was done to study the effect of re-oxidizing the surface. TE1PAV005 was oxidized in a different way—it was vented with nitrogen, partially disassembled, left for 10 minutes exposed to cleanroom air, then reassembled and pumped out. TE1PAV008 was also treated with subsequent HF (hydrofluoric acid) rinses and remeasured.

## III. RF RESULTS – EP BASELINE

$Q_0$ vs $E_{acc}$ curves were be obtained using the standard method of calibrated measurements of forward, reflected, and transmitted RF power into the cavity (for description of the technique, see [10]). Measurements were made at 2.0 K—a typical operating temperature for a niobium SRF cavity in an accelerator—and at a temperature in the 1.4-1.5 K range—close to the minimum temperature possible in the dewar, in a range where the BCS resistance is expected to be small compared to the residual resistance. In order to minimize $R_{res}$ due to trapped flux dissipation, the cavities under measurement have all been previously heat treated at high temperatures to improve flux expulsion and are all cooled in compensated magnetic field with strong thermal gradients [18]–[21]. The curves are shown in Figure 2 for the three cavities that had EP baseline before heat treatment.

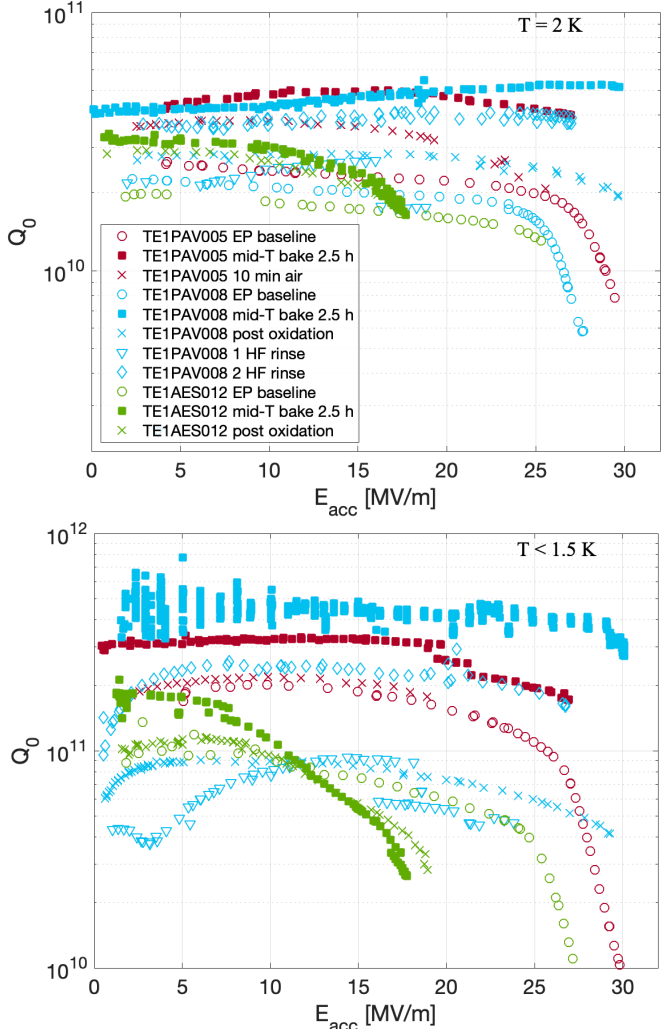

Figure 2: Quality factor versus accelerating gradient in vertical cryogenic tests of the three EP baseline cavities treated with mid-T bake. Measurements were performed at 3 stages: baseline, after mid-T bake, and after oxidation, except that no $T < 1.5$ K measurement was made on TE1PAV008 during baseline testing. Measurements were performed at a cavity temperature of 2.0 K (top) and 1.4-1.5 K (bottom).

Q vs E curves for the EP baseline cavities are fairly typical, including the observation of the high field $Q$-slope in the typical gradient range 25-30 MV/m.

After mid-T bake, extremely high quality factors were observed in TE1PAV008 and TE1PAV005. At temperatures 1.4-1.5 K (bottom of Figure 2), $Q_0$ values in the range of 3-4x10$^{11}$ were observed to gradients as high as 30 MV/m. While temperatures in the range of 2x10$^{11}$ have been reported previously [19], [22], [23], $Q_0$ above 3x10$^{11}$ at a gradient ~20 MV/m is unprecedented. At 2.0 K, the mid-T baked cavities showed an unexpected anti-$Q$-slope behavior—an increase in $Q_0$ to ~3-4x10$^{10}$ with increasing $E_{acc}$ in the range 5-15 MV/m. This behavior, which is typically associated with nitrogen-doped cavities [24], is also unprecedented in this frequency and gradient range without nitrogen doping/infusion.

After TE1PAV008 was oxidized by exposure to air and water, its $Q_0$ degraded at all fields compared to immediately after mid-T bake, but was still higher than the EP baseline at 2 K. The degradation was even smaller for TE1PAV005, which only received a 10 minute exposure to air.

After the mid-T bake and after oxidation, TE1AES012 showed Q-slope degradation not observed in other two cavities. This may be related to the leak that was observed after the mid-T bake.

The gradient limit in the baseline tests was HFQS. After mid-T bake, no HFQS was observed in either TE1PAV008 or TE1PAV005, though both cavities were limited by quench ~30 MV/m. TE1PAV005 also showed some $Q$-switch behavior in the gradient range 20-30 MV/m. Both cavities that had their oxides regrown after mid-T bake maintained very similar quench fields.

No x-rays above background were observed in any of the measurements except during some minor processing activity in TE1PAV005 in the test after 10 minute air exposure.

No errorbars are included in Figure 2 for visual clarity, but we briefly discuss measurement uncertainty here. TE1PAV005 was tested with an input coupler with $Q_{ext}$ of 9x10$^9$. This meant that for the post-mid-T bake and post-oxidation RF measurements at <1.5 K, the cavity was strongly overcoupled, increasing uncertainty. Under nominal conditions, the uncertainty in $Q_0$ and $E_{acc}$ is expected to be approximately 10%, but for $Q_0$ this much larger than $Q_{ext}$, we expect uncertainty closer to ~50%. After seeing these results, much shorter antennas were used for the tests of TE1AES012 and TE1PAV005, with $Q_{ext}$ ~5x10$^{10}$. This is expected to reduce uncertainty much closer to the 10% level.

The Q vs E curves are decomposed in Figure 3. The surface resistance in the 1.4-1.5 K range is plotted in the top of the figure, and the bottom, $R_{BCS}$ is plotted, calculated by subtracting the 1.4-1.5 K resistance from the 2.0 K resistance.

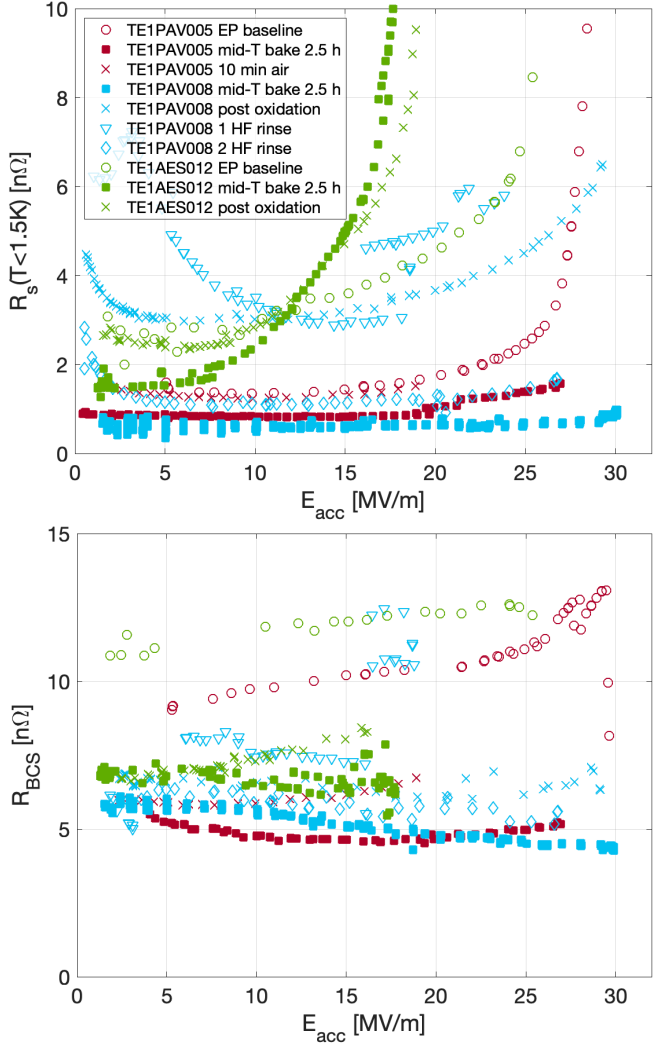

Figure 3: Decomposed surface resistance of the three cavities from Figure 2. $R_{BCS}$ (bottom) is calculated by subtracting the surface resistance at 1.4-1.5 K (top) from the surface resistance at 2.0 K.

A decrease in $R_{BCS}$ as a function of field is observed in the gradient range 5-15 MV/m. No decrease is observed in the <1.5 K curve. This suggests that the anti-Q-slope at 2.0 K originates from the BCS resistance, similar to the anti-Q-slope observed in doped cavities.

The plot of $R_s$(T<1.5 K) illustrates the extremely small surface resistance measured in TE1PAV008 and TE1PAV005. For TE1PAV005, surface resistance was measured as a function of temperature to extract the residual resistance by fitting $R_{BCS}$ to Matthis Bardeen theory [11], [25], [26]. The measurement was performed at 16 MV/m to show that the small residual resistance is achieved even at useful accelerating fields. The result is shown in Figure 4. The extrapolation to $R_{BCS}$ → 0 from Matthias Bardeen theory results in a residual resistance of just 0.63 ± 0.06 nΩ.

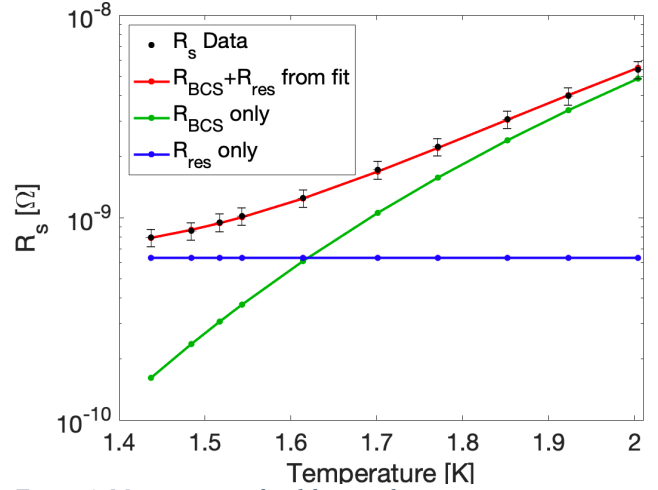

Figure 4: Measurement of and fit to surface resistance vs temperature data at 16 MV/m for TE1PAV005 after mid-T bake. The fit gives a residual resistance of just 0.63 ± 0.06 nΩ

In addition to measurements of $R_s$ vs $T$, frequency $f$ vs $T$ was measured close to $T_c$, and the combined data were used to extract material properties of the superconductor [27]. The extracted values are in Table 1.

| Cavity | $T_c$ (K) | $l$ (nm) | $\Delta/k_BT_c$ |
|---|---|---|---|
| TE1PAV005 | 8.94 ± 0.15 | 48 ± 24 | 2.07 ± 0.03 |
| TE1PAV008 | 9.04 ± 0.15 | 167 ± 84 | 2.10 ± 0.03 |
| TE1RI006 | 8.93 ± 0.15 | 62 ± 31 | 2.10 ± 0.03 |

Table 1: Critical temperature $T_c$, mean free path $l$, and reduced energy gap $\Delta/k_BT_c$ extracted from fits to measurements of $R_s$ and $f$ vs $T$ for cavities TE1PAV005, TE1PAV008, and TE1RI006 after mid-T bake.

Sensitivity was measured for TE1PAV005 after mid-T bake by measuring Q vs E after cooling slowly in a 20 mG field. The result is shown in Figure 5. The sensitivity is higher than that of an electropolished cavity, resembling that of a nitrogen-doped cavity [21].

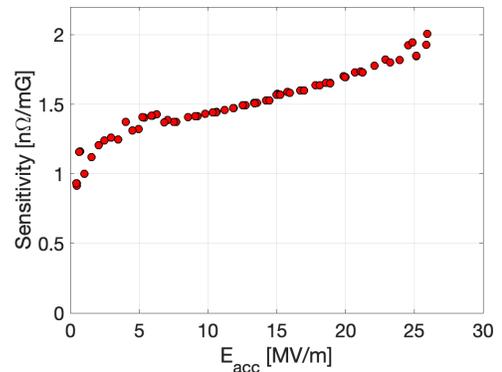

Figure 5: Sensitivity to trapped flux as a function of accelerating field for TE1PAV005 after mid-T bake, prior to venting.

## IV. RF RESULTS – 120 C AND N-DOPED BASELINE

The initial results of the mid-T bake encouraged additional studies. In addition to the three cavities that received EP baseline, three cavities were given mid-T bake that had a different baseline treatment. TE1RI006 was 3/60 N-doped (furnace treated at 800 C with 3 minutes of N-injection followed by 60 minutes of annealing) with 10 micron EP. TE1PAV011 was given EP followed by a low temperature bake at 120 C for 48 hours. TE1AES011 was given EP plus low temperature bake at 75 C for 4 hours followed by 120 C for 48 hours. The mid-T bake of TE1AES011 was slightly different. It was given an extra long 22-hour mid-T bake then the temperature was lowered to 120 C, where the cavity was vented to 25 mtorr of nitrogen, and it remained in this state for 48 hours. The $Q_0$ vs $E_{acc}$ curves of these cavities are shown in Figure 6.

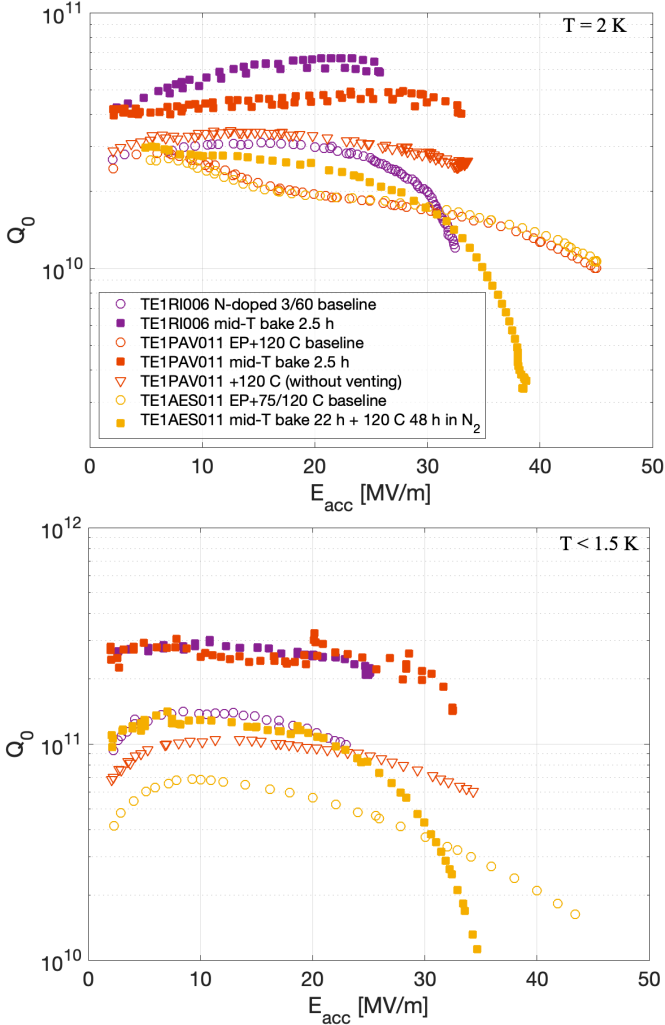

Figure 6: Quality factor versus accelerating gradient in vertical test of three cavities treated with mid-T bake after baseline treatments of N-doping or low temperature baking. TE1AES011 was given additional low temperature nitrogen treatment at 120 C. The gradient limitation in each test was quench.

Similar to the EP baseline cavities, the $Q_0$ vs $E_{acc}$ data are decomposed into residual and BCS resistance in Figure 7.

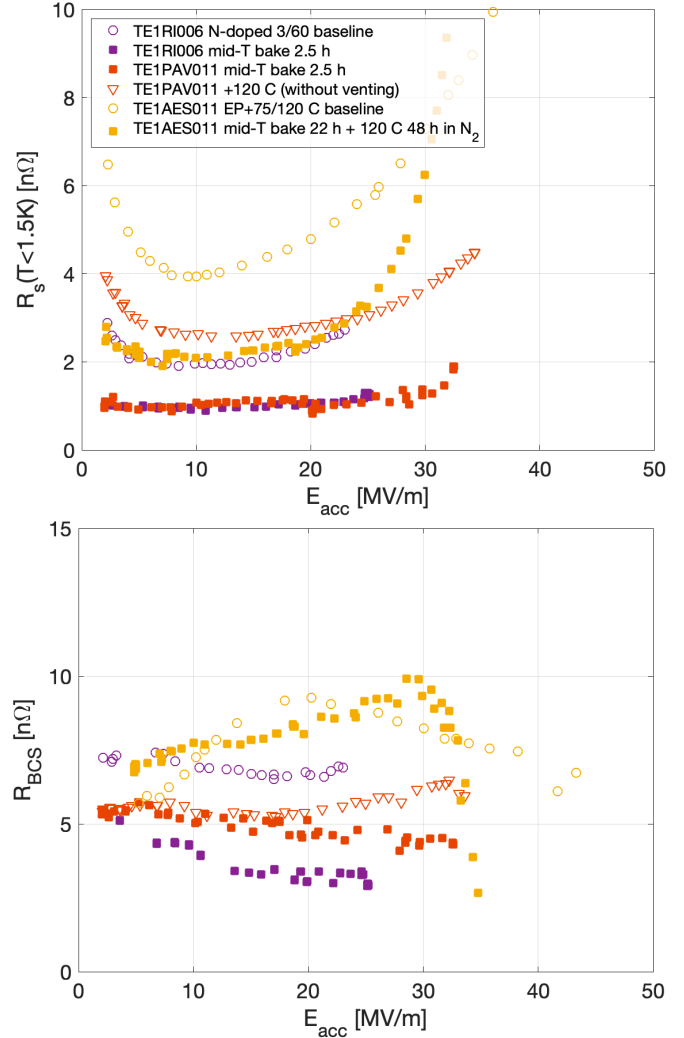

Figure 7: Decomposed surface resistance of the three cavities from Figure 6. $R_{BCS}$ (bottom) is calculated by subtracting the surface resistance at 1.4-1.5 K (top) from the surface resistance at 2.0 K.

## V. MICROSCOPIC MEASURMENTS

Microscopy of mid-T baked samples is challenging because the transfer of samples from a bake setup to a microscope must be done under vacuum in order to study the effect of the bake on the oxide. Therefore, to perform this study, electropolished niobium samples were put into a SIMS (secondary ion mass spectrometry) tool with an in-situ heating stage that could reach the appropriate temperature range. Measurements were made both on EP baseline samples, on samples after mid-T bake, and samples after exposure to air. The results are shown in Figure 8. The counts are given a point-by-point normalization to the signal collected for Nb.

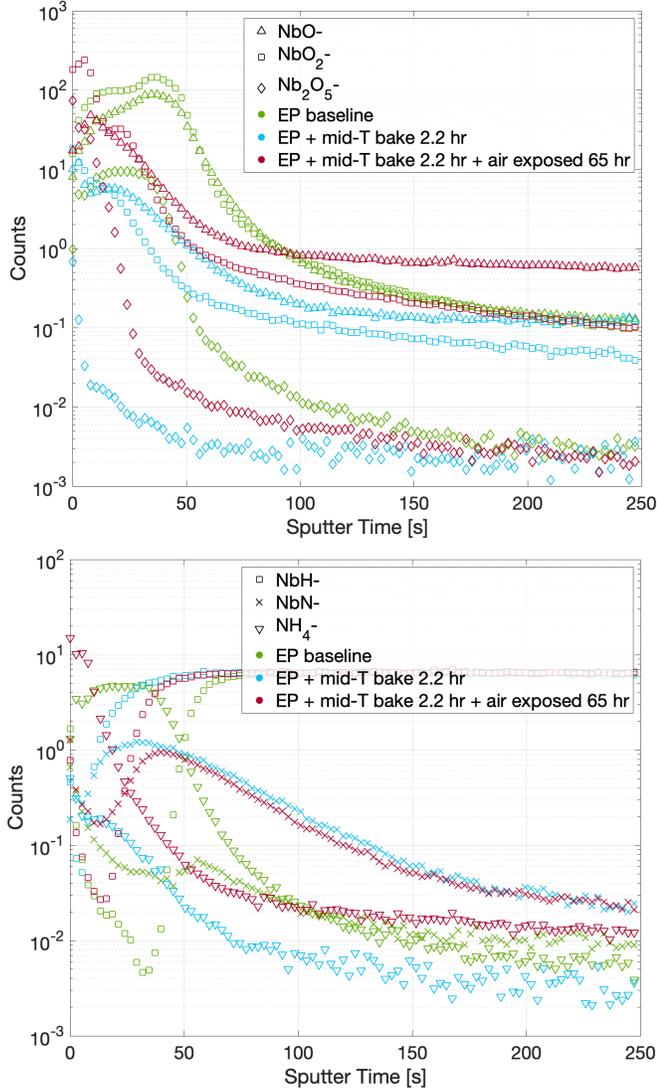

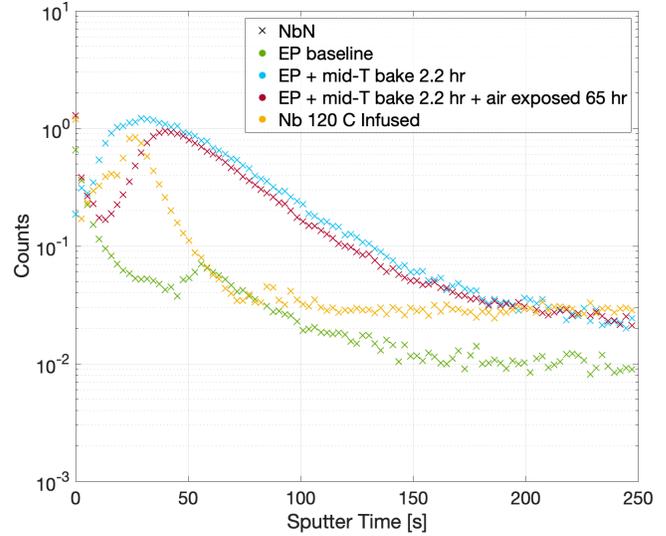

*Figure 9: Comparison of nitrogen signal in SIMS data between EP baseline, mid-T bake (with and without air exposure), and 120 C infused sample.*

*Figure 8: SIMS depth profile measurements of different species in a niobium sample that was first given EP baseline, then given mid-T bake, then exposed to air.*

The upper plot if Figure 8 plots measured intensity of Nb oxides. The $Nb_2O_5$ oxide layer is expected to be 3-5 nm thick, and it appears to sputter in 30-50 s. This gives an approximate calibration of sputter time to depth of 10 s/nm. After the mid-T bake, the $Nb_2O_5^-$ signal in the SIMS drops very quickly to the noise level, indicating a substantial reduction of the oxide layer. Some NbO and $NbO_2$ remains, but at a much lower intensity than before the mid-T bake. After exposing to air, the oxide regrows, though it appears to be somewhat thinner than the initially EP'd sample. This may be due to growing the oxide in air rather than in water.

The lower plot of Figure 8 shows intensity of $NbH^-$ and $NbN^-$ as a function of depth as a measurement of H and N interstitials in the niobium bulk. The H content is fairly similar, but the plot shows an unexpected increase in the intensity of the $NbN^-$ signal after the mid-T bake. A bump, some tens of nm deep, appears close to the surface, similar to a nitrogen infused sample. It persists after the oxide is regrown. Figure 9 compares this bump to that observed in a sample that was separately treated in a vacuum furnace with a 120 C infusion.

## VI. DISCUSSION

**Effect of Surface Condition on RF Performance**

One of the changes consistently observed was a decrease in residual resistance. Below 1.5 K, $R_s$ values were regularly observed in the range of ~1 nΩ or less. This may be related to the dissolution of the $Nb_2O_5$ layer, confirmed by SIMS measurements. A contribution by the natural oxide to residual resistance of some tenths of a nΩ would be consistent with the measurements presented here. An increase in residual resistance with oxide thickness is also consistent with measurements on anodized cavities in the relevant gradient range [28]. The reduced residual resistance may have also been due to the diffusion of impurities. For example, nitrogen doping and nitrogen infusion are also known to cause a decrease in the residual resistance in this gradient range [7], [24]. There is also a theory that suggests a contribution to residual resistance from direct photon excitation (see e.g. [29]), which may be reduced by oxide dissolution.

The very low value of residual resistance measured contributes to a limited body of knowledge around understanding of the "intrinsic" residual resistance (i.e. not caused by extrinsic factors such as trapped flux). By minimizing the effects of trapped flux (by minimizing the ambient field during cooldown and applying flux expulsion) and minimizing the effect of the oxide (by removing or significantly reducing it), we place an upper bound on the contribution of the "intrinsic" residual resistance of just 0.6 nΩ for a cavity of this type.

RF results consistently showed a decrease in BCS resistance after mid-T bake, and frequently showed anti-Q-slope. This unexpected result may be due to nitrogen interstitials, as suggested by the bump observed in the NbN signal observed by SIMS after the mid-T bake. The bump is consistent with a nitrogen infusion, though no nitrogen gas was introduced during the heat treatment in the SIMS. It is possible that the nitrogen diffused into the niobium from elsewhere, such as from surface contamination or from the bulk. Figure 9 shows some quantity of ammonia on the surface prior to heat treatment, which may have contributed to N interstitials. Other possible sources of BCS modification include O and C interstitials modifying the mean free path. The higher sensitivity and lower mean free path are consistent with measurements of nitrogen doped/infused cavities [7], [24].s

The quench field for most of the cavities after mid-T bake is in the range 25-30 MV/m. The consistency of this quench field is reminiscent of quenches in nitrogen-doped cavities, which have been observed to cluster in a narrow gradient range [30]. The only cavity with a quench field higher than this is TE1AES011, which, after the mid-T bake, was the only cavity that was backfilled to 25 mtorr with nitrogen and then received a 120 C heat treatment. This cavity reached 39 MV/m before quenching. This shows progress towards the goal of this nitrogen infusion-like treatment. Additional exploration of the parameter space (time, temperature, nitrogen pressure) may help to increase the field further. The slope that was observed is shallower than typical HFQS—additional studies of the parameters space may help to develop understanding in the cause of the slope and prevent it in future treatments.

TE1AES012 showed substantially poorer performance after mid-T bake compared to the other cavities. This may be linked to the leak during the heat treatment, allowing atmospheric gases into the cavity even with the argon purge. For example, this could have allowed undesirable oxides to form.

TE1PAV008 shows the effect of re-oxidation of the surface by exposure to water. BCS resistance stays roughly the same, though possibly slightly increased, consistent with the continued presence of nitrogen impurities in the RF layer. Residual resistance increases significantly, but it appears that this can be ameliorated by two HF rinses to return to ~2 nΩ, similar to the post-EP surface of TE1PAV005 or the N-doped surface of TE1RI006. This is consistent with the residual resistance increase observed in a 120 C baked cavity followed by subsequent HF rinsing [31].

**Previous Studies**

F. Palmer et al. previously performed studies on vacuum heat treatment to study the effect of the oxide. Palmer's experiments included firing cavities in a vacuum furnace at temperatures in the 1200-1400 C range, RF testing them without exposing to air, oxidizing the surfaces, heat treating in vacuum at temperatures ~300 C for 5-10 minutes, and then RF testing again without exposing to air. The cavities Palmer worked with were substantially higher frequency, 8.6 GHz which makes comparison of results complicated but still interesting. Palmer's results show the lowest residual resistance after 1200-1400 C firing and before exposing to oxygen. $R_{res}$ was found to have a similar small value after oxidation and heat treatment to 350 C for 10 minutes, but heat treatment at 300 C or 325 C resulted in significant increases in residual. This suggests a strong temperature dependence, and may also imply some dependence on time at these temperatures. The residual value after 350 C heat treatment was smaller than the value after initial bulk chemistry, in agreement with the results presented here. Interestingly, Palmer also shows a decrease in BCS resistance after 325 C heat treatment, also consistent with the results presented here. They propose that diffusion of oxygen causes a change in mean free path that in turn brings the BCS resistance closer to a minimum. The role of nitrogen is not discussed [13], [32]–[34].

G. Eremeev also performed previous studies of the effect of heat treatment in vacuum at ~400 C to study the effect of the oxide on high field Q-slope. Because of the focus of these experiments, it is difficult to compare results: measurements were performed only at 1.5 K, so BCS resistance at 2 K cannot be extracted; there are limited comparisons to data before heat treatment at ~400 C in vacuum; and the residual resistance is relatively high even before heat treatment ($Q_0$~3x10$^{10}$ at low fields at 1.5 K) [14].

**Outlook: Future Improvement and Applications**

Future work will focus on parameter space exploration with the goal of improving performance. Possible avenues to explore will include trying to optimize the nitrogen profile using feedback between SIMS studies, light removal post mid-T bake, and cavity testing, as well as studying if the oxide can be more fully dissociated by the heat treatment (consider the near-surface NbO and NbO$_2$ signals in Figure 8). Treatments with the best performance will be evaluated for reliability, ideally including studies at other labs. Additionally, cavities at other frequencies and multicells will be treated with the mid-T bake. In addition to possible applications, studies of this treatment may help to improve understanding of "intrinsic" residual resistance and of the $Q_0$ improvement brought about by nitrogen doping and infusion.

For application in accelerators, it would be very challenging to use cavities in accelerators after mid-T bake but before exposure to air. It seems more likely that oxidized surfaces would have to be used in accelerator applications. The performance of TE1PAV005 after 2 HF rinses shows that this is a promising possibility. With $Q_0$ of 3.9×10$^{10}$ at 27 MV/m, this cavity would significantly exceed the planned specifications for LCLS-II HE, making it a possible treatment to study for future high duty factor applications.

Additionally, there are non-accelerator applications requiring extremely high $Q_0$ electromagnetic resonators where exposure to air after heat treatment is not needed or else where only a very brief exposure is needed (duration similar to that used for TE1PAV005). These include applications such as quantum computing [28], [35] and quantum sensors.

## VII. CONCLUSIONS

In this study, several 1.3 GHz single cell cavities were first assembled in a cleanroom to RF hardware and put under vacuum, then given a mid-T bake: a heat treatment for several hours at a temperature of 250-400 C, which is expected to be sufficient to remove or significantly reduce the oxide. The mid-T bake was found to 1) significantly decrease residual resistance, and 2) significantly reduce BCS resistance at 2 K relative to an EP'd cavity. $Q_0$ values of 3-4×10$^{11}$ were measured <1.5 K and 20 MV/m, higher than has been reported previously in the literature for such conditions, and a residual resistance of just 0.63 ± 0.06 nΩ was measured at 16 MV/m. After exposing to the cavities to air and water, the $Q_0$ degraded somewhat but was still ~3×10$^{10}$ at 2 K, similar to N-doped cavities. The observation of anti-$Q$-slope in the BCS resistance, the increased sensitivity to trapped flux, and the nitrogen observed in SIMS are all consistent with the $Q_0$ of the cavity being enhanced due to the presence of nitrogen interstitials. HF rinse studies so far show some potential to improve $Q_0$ at 2 K to nearly 4×10$^{10}$ at 27 MV/m. The quench field was found to be fairly consistently in the 25-30 MV/m range after 2.5 hours of mid-T bake, but the addition of a 120 C step for 48 hours with 25 mtorr of nitrogen, similar to N-infusion was found to increase the quench field to 39 MV/m. Future studies will focus on continued optimization of the process as well as exploration of other cavity frequencies and multicells.

## ACKNOWLEDGEMENTS

The authors would like to thank B. Tennis for preparation of the mid-T baking setup as well as the Fermilab SRF processing and VTS teams. This work was supported by the United States Department of Energy, Offices of High Energy Physics and Basic Energy Sciences under Contract DE-AC05-06OR23177.